\documentclass[11pt,a4paper]{article}
\usepackage[T1]{fontenc}
\usepackage[utf8]{inputenc}
\usepackage{authblk}
\usepackage{amsmath,amssymb,epsfig,cite,dsfont}
\usepackage{array,multirow}
\usepackage {color}
\usepackage[toc,page]{appendix}
\numberwithin{equation}{section}

\usepackage[colorlinks=true
,urlcolor=blue
,anchorcolor=blue
,citecolor=blue
,filecolor=blue
,linkcolor=blue
,menucolor=blue
,pagecolor=blue
,linktocpage=true
,pdfproducer=medialab
,pdfa=true
]{hyperref}

\advance\hoffset by -1.1truecm
\advance\voffset by -1.0truecm
\newcommand{\be}{\begin{equation}}
\newcommand{\ee}{\end{equation}}
\newcommand{\bea}{\begin{eqnarray}}
\newcommand{\eea}{\end{eqnarray}}

\usepackage{mathrsfs}
\usepackage{authblk}

\numberwithin{equation}{section}
\usepackage{float}
\restylefloat{figure}
\newcounter{appendice}

\begin{document}

\title{
\vspace{2.2cm}\begin{flushleft} \bf{ Matrix Model of QCD: Edge Localized Glue Balls  and Phase Transition
\linethickness{.05cm}\line(1,0){433}
}\end{flushleft}}
\author[1]{Nirmalendu~Acharyya \thanks{nacharyy@ulb.ac.be}}
\author[2,3]{A.~P.~Balachandran \thanks{balachandran38@gmail.com}}
\affil[1]{\small Optique Nonlin\'eaire Th\'eorique, Universit\'e Libre de Bruxelles (U.L.B.), CP 231, Belgium}
\affil[2]{\small Physics Department, Syracuse University, Syracuse, New York 13244-1130, U.S.A.}
\affil[3]{\small Institute of Mathematical Sciences, C.I.T Campus, Chennai, TN 600113, India}
\date{}

\renewcommand\Authands{ and }

\maketitle

\begin{abstract}
{ In a matrix model of pure $SU(2)$ Yang-Mills theory, boundaries emerge in the space of $\textrm{Mat}_{3}(\mathbb{R})$ and the Hamiltonian requires boundary conditions. We show the existence of edge localized glueball states which can have negative energies. These edge levels can be lifted to positive energies if  the gluons acquire a London-like mass.  This suggests a new phase of QCD with an incompressible bulk. }

\end{abstract}

\section{Introduction}

{ Quantum chromodynamics or QCD describing strong interactions is an interacting non-Abelian gauge theory.} Non-abelian gauge theories of high energy physics are based on compact gauge groups. They generally contain a Lie group like $SU(3)$ or $SU(2)$ which modulo their discrete centres are simple.  { The self-coupling of the gauge field in such a gauge theory is expected to lead to bound states called glueballs. In the presence of matter (say, quarks), these particle excitations interact with hadrons. It is for such reasons that glueballs remain an interesting topic of investigation, despite the fact that they have eluded experimental verification till now.}

{ In a non-Abelian gauge theory, it is impossible to do a global gauge fixing. }
In particular, Gribov \cite{gribov} showed that in all such theories, the Coulomb gauge does not fully eliminate the gauge freedom: there are gauge related copies of the connection in this gauge. It was later proved { rigorously} by Singer \cite{singer} and by Narasimhan and Ramadas \cite{narasimhan} that there exists in fact no condition to eliminate the gauge ambiguities, the gauge bundle on the configuration space being twisted.

Narasimhan and Ramadas, in their work on $SU(2)$, reduced the considerations to a family of connections parametrised by $3 \time 3$ real matrices. The essential topological complexities of exact {pure Yang-Mills theory} are already captured by this model. Here too the appropriate $SU(2)/\mathbb{Z}_2 = SO(3)$ bundle is twisted, the twist being inherited from the full {pure Yang-Mills theory}.

The work of Narasimhan and Ramadas can be extended to $SU(3)$ and other non-abelian groups. That is because it is based on Maurer-Cartan forms which have a certain universal character.  {Recently, in \cite{Balachandran:2014voa,mm_bsa},  a matrix model of the $SU(N)$ Yang-Mills theory was proposed which successfully captures the non-trivial nature of the gauge bundle. There, the Hamiltonian formalism for these matrices as configuration spaces was deduced from the full Yang-Mills theory. The matrix model is constructed by compactifying the spatial $\mathbb{R}^3$ to $S^3$. The Maurer-Cartan form of $SU(N)$ is  pulled back on the $S^3$ to obtain a particular subspace of the space of all gauge fields. In this subspace, the gauge fields are $3 \times (N^2-1)$ real matrices and the result  is the (0+1)-dimensional matrix model of $SU(N)$.

In\cite{Balachandran:2014voa,mm_bsa}, the Hamiltonian formalism for these matrices as configuration spaces was deduced. The colorless eigenstates of the Hamiltonian are interpreted as ``glueballs'' and it is shown that the glueball spectrum for the $SU(2)$ gauge group has a mass gap. The presence of this gap is often regarded as a signal for confinement. There,  QCD $\theta-$angle is also discussed and the   Dirac operator is constructed. In a numerical study \cite{Acharyya:2016fcn},  the authors obtained the estimates for glueball masses in the $SU(3)$ matrix model  and found an excellent agreement with those obtained from lattice QCD simulations \cite{Morningstar}, despite the numerics being far simpler and less time consuming in the matrix model.  This indicates that the matrix model might emerge as an efficient tool for QCD computations with fair accuracy. This motivates us to further investigate various other aspects of the matrix model in detail, as they can carry useful implications about the full pure Yang-Mills theory. }

%In two previous papers by Balachandran, Queiroz and Vaidya \cite{Balachandran:2014voa,mm_bsa}, the Hamiltonian formalism for these matrices as configuration spaces was deduced from QCD. The glueball spectrum for $SU(2)$ gauge group was found to have a mass gap. The presence of this gap is often regarded as a signal for confinement. They also discussed QCD $\theta-$angle, constructed the Dirac operator and gave estimates for glueball masses using variational methods. 

In this paper, we { study} certain ``singular''  boundaries of the $3\times3 $ matrix model of $SU(2)$ Yang-Mills and the special states localised at these boundaries.  Such boundary states exist also for $SU(3)$ and other gauge groups, being a reflection of states localised at degenerate connections in exact QCD.

In the space  $\textrm{Mat}_{3}(\mathbb{R})$ of $3 \times 3$ real matrices, boundaries and stratification emerge as follows. A matrix  $M \in \textrm{Mat}_{3}(\mathbb{R}) $ has the singular value decomposition
\begin{eqnarray} \label{eqn3.1}
&&M = L D R^T,  \quad\quad L,R \in SO(3) \\
&&D=\left(\begin{array}{ccc}
a_1&0&0\\
0&a_2& 0 \\
0&0&a_3
\end{array}\right), \quad\quad a_1 \geq a_2 \geq a_3 \geq 0. 
\label{eqn3.2}
\end{eqnarray}
When all the $a_i$'s are unequal,  $a_i \neq a_j$ if $i\neq j$, we get the open and dense stratum. At the boundaries, either a pair of $a_i$ or all $a_i$ are equal. 

The boundaries $\partial \mathcal{D}$ of the spatial manifold $\mathcal{D}$ have physical consequences.  The Laplace and Dirac operators are subject to boundary conditions at $\partial \mathcal{D}$  for self-adjointness. The latter can induce anomalies \cite{esteve}. They can also create edge-localised states \cite{Asorey1} which are of particular interest for topological insulators as discussed previously \cite{Asorey2}. For the Dirac operator, when the boundary conditions are of Atiyah-Patodi-Singer type, they lead to the $\eta$-invariant which has an impact on axial anomaly \cite{ninomiya}. Such boundary conditions can also make or break supersymmetry or BRST invariance \cite{Acharyya1, Acharyya2}.

These known results are the incentives to study the boundaries of $\textrm{Mat}_3(\mathbb{R})$. As for spatial manifolds with boundaries, here too the Hamiltonian requires boundary conditions. {Considering various boundary conditions, we explore the possibility of ``edge''   localised glueball states (localized near the boundary associated with $\textrm{Mat}_3(\mathbb{R})$)}. They are expected to be novel glueball states {and might imply the existence of new phases of QCD}. This work focuses on these aspects. It can be extended to $\textrm{Mat}_8(\mathbb{R})$ which is appropriate for colour $SU(3)$.  {As shown in \cite{Acharyya:2016fcn}, the glueball spectrum of the matrix model matches excellently with the physical masses predicted by lattice QCD. Similarly, these edge states might also be present in the full pure Yang-Mills theory.}

{ We here confirm first that there do exist such edge states. The energy of these edge states can be negative. Such states are physical only if the gluons  acquire mass.  This suggests the possibility of new phases in which the  gluons becomes massive just as the photon acquires a London mass in a superconductor. When matter fields are coupled to the matrix model, similar edge states emerge naturally \cite{Pandey:2016hat}. Here, we demonstrate that the emergence of such edge states is due to the presence of nontrivial boundary conditions on  $\textrm{Mat}_3(\mathbb{R})$.   Further these ``superconducting'' phases share features with earlier models of quark-gluon plasma \cite{sanatan1, sanatan2}. }

The first step in the analysis is the partial wave decomposition of wave functions $\Psi: \textrm{Mat}_3 (\mathbb{R}) \rightarrow \mathbb{C}$ with regard to the  two $SO(3)$'s appearing in (\ref{eqn3.1}). The Laplacian for the matrix model then separates as shown by Iwai \cite{iwai}. We discuss this in section 2 where we also clarify the meaning of the transformation $M \rightarrow L^\prime M R^{\prime T}$,  $L^\prime, R^\prime \in SO(3)$ which commutes with the Laplacian.  In this manner, we arrive at the $SO(3)_L \times SO(3)_R$ invariant S-wave sector of glueballs. 

The eigenvalue problem is singular at the boundaries where two or more $a_i$ becomes equal.  It is of the same kind as the singularity at $r=0$ of radial eigenvalue problem on $\mathbb{R}^d$ ($d$=dimension).  In the latter, as is known, it appears in the volume form $r^{(d-1)} d r d\Omega_{S_{(d-1)}}$ which becomes zero at $r=0$. We can {transfer} the $r^{(d-1)}$ factor to the Hamiltonian. Then the new volume form $d r d\Omega_{S_{(d-1)}}$  is well behaved at the origin, while the transformed radial Laplacian 
\begin{equation}\label{eqn_1.3}
{ -  \partial_r^2   + \frac{ (d-3)(d-1)}{4 r^2}  + \frac{l(l+d-2)}{r^2}}
\end{equation}
has acquired the singular potential $\frac{ (d-3)(d-1)}{4 r^2} $ for $d \neq 1,3$.   For all other  values of $d$, the singularity at $r=0$ calls for special boundary conditions which can be found using Weyl's ``limit point-limit circle'' theorems \cite{coddington_levinson}. {Notice that for $1< d<3$, the potential is attractive whereas for all other values, it is repulsive. } In a similar way, in our glueball problem, a potential with singularities of the form $\prod_{i > j} (a_i^2 - a_j^2)^{-1}$  appears. Fortunately, they are amenable to Weyl's approach. These matters are discussed in section 3 where we also bring the eigenvalue problem to a stage which can be treated by variational methods.

Section 4 reports on the variational calculation. The singularity at the boundary $\partial \textrm{Mat}_3(\mathbb{R})$ is of the ``limit circle'' type so that the self-adjoint extensions are characterised by the phases $e^{i \theta}$. The Dirichlet boundary condition has $e^{i \theta} =-1$, while the Neumann boundary condition has $e^{i\theta} =1$. For the Robin boundary conditions which are near Dirichlet, just as the spatial boundary, edge localised glueball states exist. { For certain choice of the boundary condition, these states has positive energy, while some leads to negative energy. In section 5, we give the interpretation of the negative energy states in terms of a new QCD phase with an incompressible bulk, in  close analogy to superconductivity on a spatial domain $\mathcal{D}$ \cite{Asorey1}. In the latter, there are localised low-lying states at the boundary $\partial \mathcal{D}$, whereas the bulk states are gapped. } In the new QCD phase, the gluons are massive. Such masses can be generated when matter fields are coupled to the matrix model \cite{Pandey:2016hat}.

In the section 6, we highlight certain observations of Iwai \cite{iwai}\footnote{Their significance was pointed out to us by Sachindeo Vaidya}. Namely, the Hamiltonian does not have a divergent centrifugal barrier term near the boundaries when the wave functions transform non-trivially under $SO(3)_L$ or $SO(3)_R$. This is in striking contrast to the Laplacian on $\mathbb{R}^d$ which {\textit{does}} have a centrifugal potential for non-zero angular momentum, that is for non-singlet $SO(d)$ representations. That suggests that edge states exist regardless of $SO(3)_L$ or $SO(3)_R$ angular momentum. The QCD potential from angular momentum and colour excitations does depend on these excitations and change with $SO(3)_{L,R}$ representations (although it is finite when the boundary is approached) so that the glueball excitations need not be degenerate. 

%The work on the $SU(3)$ model is in progress \cite{arshad}.  

In a different project \cite{Acharyya:2016fcn}, the glueball masses in the same matrix model have been estimated using the harmonic oscillator eigenstates. Low lying glueball spectra  obtained there are remarkably similar to the ones  from lattice QCD.

\section{The Hamiltonian and its Partial Wave Reduction}

The origin of the matrix model for QCD comes from the well-known  ``Gribov ambiguity''  \cite{gribov,singer}. We explain how that is so in this section focusing on the SU(2) gauge group. We consider $SU(2)$ and not $SU(3)$ as  our numerical work has been on $SU(2)$. Theoretical considerations on $SU(3)$ can be found in \cite{Balachandran:2014voa,mm_bsa}.

The ``Gribov ambiguity'' can be summarised in the statement that the gauge bundle in any non-abelian gauge theory which involves a compact semi-simple Lie group is twisted. Therefore, there is no global gauge fixing condition in any such theory. 

The full space of connections on $\mathbb{R}^d$ in any gauge theory is infinite dimensional. Narsimhan and Ramadas  \cite{narasimhan} proved that the exact gauge theory twist is reflected in the following finite-dimensional submanifold of connections parametrised by matrices:
\begin{eqnarray}
\Omega = \textrm{Tr}\left[ \frac{\tau_a}{2}  u^{-1} du\right] M_{ab} \tau_b \label{mc1}
\end{eqnarray}
Here we consider spatial dimension 3 and  $SU(2)$ gauge group and $\tau_a$'s are Pauli matrices. $M$ is a real $3 \times 3$ matrix and $u$ is given by the Skyrme ansatz \cite{bal_book}: 
\begin{eqnarray}
u(\vec{x}) = 
\cos \theta (r) + i \tau_i \hat{x}_i \sin \theta(r), \quad\quad 
\theta(0) =\pi, \quad\quad \theta (\infty) = 0, \quad\quad \vec{x} \in \mathbb{R}^3, \quad\quad r=| \vec{x}|, \label{eqn2.2}
\end{eqnarray}
$\theta$ being a monotonic function of $r$.

There are two group actions of interest on $M$: 
\begin{enumerate}
\item The first comes from the colour $SU(2)$ transformation
\begin{equation}
\Omega \rightarrow g \Omega g^{-1}, \quad\quad g \in SU(2).  
\end{equation}
Since 
\begin{eqnarray}
g \tau_b g^{-1} = \tau_c Ad g_{cb} 
\end{eqnarray}
where $g \rightarrow Ad g$ is the $3 \times 3$ adjoint representation of $SU(2)$, the transformations 
\begin{eqnarray}
M \rightarrow M Ad g^T 
\end{eqnarray}
are $SU(2)$ colour transformations. Observables are all colour singlets. 

Only global colour acts on $\Omega$: it is partially ``gauge fixed'' to eliminate space-time dependent transformations. 

\item  Under the transformation 
\begin{eqnarray}
u \rightarrow us, \quad\quad s \in SU(2),
\end{eqnarray}
we have 
\begin{eqnarray}
u^{-1}  du  \rightarrow s^{-1} (u^{-1} du) s.
\end{eqnarray}
Or since
\begin{eqnarray}
s \tau_b s^{-1} = \tau_c Ad s_{cb}, \quad\quad Ad s \in SO(3), 
\end{eqnarray}
this gives the transformation
\begin{eqnarray}\label{eqn2.10}
M \rightarrow Ad s M. 
\end{eqnarray}
Now $s^{-1} (u^{-1} du) s$ is also achieved by the transformation
\begin{equation}
u \rightarrow s^{-1} u s
\end{equation}
and that, as (\ref{eqn2.2}) shows, is a spatial rotation. Hence, (\ref{eqn2.10}) corresponds to spatial rotation. 

{ In brief, the matrix model is constructed by compactifying the spatial $\mathbb{R}^3$ to $S^3$ of radius $R$   and pulling back the Maurer-Cartan form on $SU(N)$ to obtain a particular subspace of the space of all gauge fields. In this subspace, the gauge fields are $3 \times (N^2 -1)$ real matrices, yielding a (0 + 1)-dimensional matrix model of SU(N) Yang-Mills theory. }

\quad We use \cite{Balachandran:2014voa, mm_bsa} for the matrix model Hamiltonian and the transformation properties of the states. The Hamiltonian is invariant under colour $SU(2)$ and spatial rotations. \\
\mbox{} \\
\textit{The Hamiltonian:}
\quad The exact pure Yang-Mills action is 
\begin{eqnarray}
\begin{array}{ll}
S_{QCD} = -\displaystyle\frac{1}{2 g^2}\int d^4 x F_{\mu\nu} (x) F^{\mu\nu} (x),  \quad\quad 
F_{\mu\nu} = \partial_\mu  A_\nu - \partial_\nu A_\mu + [A_\mu, A_\nu].
\end{array}
\end{eqnarray}
It gives the gluon Hamiltonian
\begin{eqnarray}\label{Ham1}
H_{QCD} = \frac{1}{2} \int d^3 x \textrm{Tr} \left[g^2 E_i E_i - \frac{1}{g^2} F_{ij}^2 \right],
\end{eqnarray}
where the electric field $E_i$ is conjugate to $A_i$.

The matrix model Hamiltonian follows from (\ref{Ham1}). We introduce 
\begin{equation}
E_{i\alpha} = -i \frac{\partial \,\,}{\partial M_{i\alpha} }
\end{equation}
as conjugate operators to $M_{i\alpha}$ and write the matrix model Hamiltonian
\begin{eqnarray}
\begin{array}{llll}
&&H =- \frac{1}{R} \left[ \displaystyle \frac{g^2}{2} \sum_{i, \alpha} \frac{\partial^2 \,\,}{\partial M_{i \alpha}^2}-V(M)\right], \quad\quad V(M) =- \displaystyle\frac{1}{2g^2}\textrm{Tr} F_{ij}^2.
\end{array}
\end{eqnarray}
{ $R$ is the radius of the $S^3$}. 

{The above  Hamiltonian only takes into account the classical zero-mode sector of the full field theory. To account for the contribution from the zero-point energy of all the higher, spatially dependent modes in  the full quantum field theory, we  need to add a constant $C(R)$ to the above Hamiltonian. }
\begin{eqnarray}
\begin{array}{llll}
&&H =- \frac{1}{R} \left[ \displaystyle \frac{g^2}{2} \sum_{i, \alpha} \frac{\partial^2 \,\,}{\partial M_{i \alpha}^2}-V(M) + C(R)\right].
\end{array}
\end{eqnarray}
{The R dependence comes from the fact that C(R) is the renormalized total zero-point energy (see for example \cite{degrand}). The numerical values of $R$ and $C(R)$ can be obtained phenomenologically as described in \cite{Acharyya:2016fcn}. }

The curvature $F_{ij}$ is
\begin{eqnarray}
\begin{array}{lll}
&&\displaystyle F_{ij} = ( d\Omega + \Omega \wedge \Omega ) (i X_i, i X_j),\\
&& \displaystyle X_i =\textrm{angular momentum generators. }
\end{array}
\end{eqnarray}
Since the Skyrme ansatz effectively works on $S^3$, $X_i$'s replace the spatial translations in $F_{ij}$. This curvature is computed in \cite{Balachandran:2014voa, mm_bsa}: 
\begin{eqnarray}
F_{ij} = i \epsilon_{ijk} M_{k\alpha} \frac{\tau_\alpha}{2} - i \epsilon_{\alpha\beta\gamma} M_{i \alpha} M_{j\beta} \frac{\tau_\gamma}{2}, \quad\quad i=1,2,3, \quad \alpha = 1,2,3. 
\end{eqnarray}
In this way, we get
\begin{eqnarray}
V(M) = \frac{1}{2g^2} [{ M_{ia} M_{ia}} - \epsilon_{ijk} \epsilon_{\alpha\beta \gamma} M_{i \alpha} M_{j\beta} M_{k\gamma} +\frac{1}{2} \epsilon_{\alpha_1 \beta_1\gamma} \epsilon_{\alpha_2 \beta_2 \gamma} M_{i \alpha_1} M_{j \beta_1} M_{i\alpha_2} M_{j \beta_2}.
\end{eqnarray}

The scalar product of the functions $\Psi$ for the Hilbert space on which $H$ operates is
\begin{equation}
(\Psi_1, \Psi_2) = \int \prod_{i,\alpha} dM_{i\alpha} \bar{\Psi}_1(M) \Psi_2(M)
\end{equation}
\end{enumerate}

 Using the singular value decomposition
\begin{eqnarray}
M= R A S^T, \quad\quad A\equiv \left(\begin{array}{ccc}
a_1&0&0\\
0& a_2&0\\
0&0& a_3
\end{array}\right), \quad\quad { a_1 \geq a_2 \geq a_3 \geq 0},
\end{eqnarray}
we get the simple expression
\begin{eqnarray}\label{pot} 
V(M)= \frac{1}{2 g^2}\left[ (a_1^2 +a_2^2+a_3^2) -6 a_1 a_2 a_2 + (a_1^2 a_2^2 +a_2^2 a_3^2 +a_3^2 a_1^2 )\right]
\end{eqnarray}

The next step for $SU(2)$ or more precisely SO(3), is separation of variables for $SO(3)_L \times SO(3)_R$ which act on the left and right of $M$ respectively. This work has been done by Iwai \cite{iwai}. As he shows, if $d\Omega_{L,R} $ are the $SO(3)$-invariant volume forms of $SO(3)_{L,R}$, then 
\begin{eqnarray}\begin{array}{lll}
&&\prod_{i,\alpha} { dM_{i\alpha}} = \phi(a) \prod_{i} da_i d \Omega_L d\Omega_R, \quad\quad  \phi (a) = (a_1^2-a_2^2)(a_2^2-a_3^2)(a_1^2-a_3^2) \geq 0, \\ \\
&& d \Omega_{L,R} = SO(3)_{L,R} \textrm{ invariant volume forms.}
\end{array}
\end{eqnarray}

Also, when acting on $SO(3)_{L,R}$ singlet wave functions, which are our subject of numerical investigations, the Laplacian $- \sum \frac{\partial^2\,\,} {\partial M_{i\alpha}^2}$ reduces to 
\begin{eqnarray}  \label{eqn_delta} {
- \sum \frac{\partial^2\,\,} {\partial M_{i\alpha}^2}\rightarrow {\Delta}} &=& {- \left[\frac{\partial^2\,\,} {\partial a_{1}^2} + 2a_1 \left(\frac{1}{a_1^2 - a_2^2}+ \frac{1}{a_1^2-a_3^2}\right)\frac{\partial\,\,} {\partial a_{1}}\right.} \nonumber \\
&& \ \ \, { \left.+ \frac{\partial^2\,\,} {\partial a_{1}^2} + 2a_2 \left(\frac{1}{a_2^2 - a_1^2} + \frac{1}{a_2^2-a_3^2}\right) \frac{\partial\,\,} {\partial a_{2}}\right.} \nonumber \\
&&  \ \ \, { \left. + \frac{\partial^2\,\,} {\partial a_{3}^2} + 2a_3 \left(\frac{1}{a_3^2 - a_1^2} + \frac{1}{a_3^2-a_2^2}\right)\frac{\partial\,\,} {\partial a_{3}}\right].}
\end{eqnarray}

Since, $d\Omega_{L,R}$ only  supply overall factors in the scalar product of singlets, we will  ignore them. 

It is convenient to change the volume form to $\prod da_i$ by changing $\Delta$ to
\begin{eqnarray}\label{H0}
\frac{2R}{g^2} \hat{H}_0 = \sqrt{\phi} \Delta \sqrt{\phi}
\end{eqnarray}
and hence $H$ to 
\begin{equation}
\hat{H} = \hat{H}_0 +V(M). 
\end{equation}

The expression for $\frac{2R}{g^2} \hat{H}_0$ is
\begin{eqnarray}
&&\frac{2R}{g^2} \hat{H}_0 = - \sum_{i=1}^{3} \frac{\partial^2 \,\,}{\partial a_i^2} + U(a) \label{eqn2.17} \\
&& { U(a)= \frac{1}{2} \left(\frac{1}{\phi}\frac{\partial^2\phi}{\partial a_i^2} \right)(a) -  \frac{1}{4} \left(\frac{1}{\phi}\frac{\partial \phi}{\partial a_i}\right)^2(a), }
\end{eqnarray}
%where primes denote derivatives with respect to $a_i$. 
In more explicit form,
\begin{eqnarray}
 U(a)&=&\frac12 \left({\frac{1}{(a_1-a_2)^2}+\frac{1}{(a_1+a_2)^2}+\frac{1}{(a_1-a_3)^2}+\frac{1}{(a_2-a_3)^2}+\frac{1}{(a_1+a_3)^2}+\frac{1}{(a_2+a_3)^2}}\right)\nonumber\\
 &=&\displaystyle{\frac{a_1^2+a_2^2}{(a_1^2-a_2^2)^2}+\frac{a_1^2+a_3^2}{(a_1^2-a_3^2)^2}+\frac{a_2^2+a_3^2}{(a_2^2-a_3^2)^2}. }
\end{eqnarray}

\section{On Boundary Conditions} 

The new potential $U(a)$ is singular as $a_i \rightarrow a_j$ { due to the stratified structure of the matrix orbit space.  The bulk corresponds to the orbits of irreducible gauge fields, whereas the boundary contains the orbits of reducible gauge fields. 
{ In the pure gauge theory,  there are natural boundary conditions for the quantum Hamiltonian. When the gauge theory is coupled to matter fields, some more general conditions can be considered see e.g. \cite{GK:1991,FG:1994, AFLL:1995, AFLL:1997, AS:2012,AS:2013,AS:2014,AS:2015}.  Therefore, one needs to consider the possibility of more general boundary conditions to treat these singularities.}
%{\color{red} Although in the absence of matter, there are natural boundary conditions for the quantum Hamiltonian, in the presence of matter insertions some more general conditions can be considered see e.g. \cite{GK:1991,FG:1994, AFLL:1995, AFLL:1997, AS:2012,AS:2013,AS:2014,AS:2015}. Therefore, one needs to consider the possibility of more general boundary conditions to treat these singularities. }}{ \color{blue} \bf Comment for Manolo: We have  no matter.  Should be clarified.}

The analogous problem for boundary conditions  in one variable was treated by Weyl \cite{weyl} and goes by the name of  ``limit point, limit circle'' theorems \cite{reed_simon,coddington_levinson}. The generalisation of Weyl's approach to several variables is due to Harishchandra and is described by Knapp \cite{knapp}\footnote{We thank  Professor M.S. Narasimhan who helped us in understanding this work and for these references.}.

Fortunately, because of certain simplicities, we can treat the domain of self-adjointness of (\ref{eqn2.17}) without the full machinery in Knapp. The approach we follow is due to \cite{Asorey:2004kk, AS:2012,AS:2013,AS:2014,AS:2015}. 

%First, one notices that since $\hat{H}_0$ has real coefficients, its deficiency indices are equal and therefore admits one or more domains of self-adjointness. 

The general method here for finding all boundary conditions is as follows.  
%Let $\Lambda >>1$ be a large constant cutoff. Consider the region of $a_i$'s where 
%\begin{equation}\label{eqn3.11}
% \phi(a) \geq \Lambda. 
%\end{equation}
%The operator $\hat{H}_0$ acts on smooth functions in this interval.  
Let us consider the   asymptotic zero modes $\Psi_{azm}$  of $ \hat{H}_0$ that are square integrable in a neighborhood of the singularities of the effective  potential, that is % $\phi(a) \geq \Lambda$ ,  i.e.
\begin{eqnarray}
\hat{H}_0 \Psi_{azm} = \mathcal{O}\left (\frac1{\Lambda^2} \right), 
\end{eqnarray}
with
\begin{eqnarray}
&&  \int_{ \phi(a)\leq \Lambda } \prod_i da_i |\Psi_{azm}(a)|^2  < \infty,
\end{eqnarray}
$\Lambda \ll 1$ being an arbitrary small cutoff.
{In general there is an infinity of such  asymptotic zero-modes  \cite {AS:2013}}. However, they can be parametrized
 by separation of variables in a way similar to what is done in the one dimensional case (\ref{eqn_1.3}). We can choose a
system of coordinates on the surface $\phi(a)=\Lambda$ and one extra {\it radial} coordinate given by ${\phi(a)^{\frac13}}$.
The Hamiltonian $\hat{H}_0$ then splits into two parts: one {\it radial} term $H_\phi$ and one {\it angular} term $H_\Lambda$. The asymptotic
zero modes can then be split according to the different eigenvalues $\lambda_n\geq 0$ of angular part. 
% For each non-vanishing eigenvalue of  $\Delta_\Lambda$ there is a unique asymptotic zero mode of $\hat{H}_0$, whereas for the vanishing eigenvalue $\lambda_0=0$,  there are the independent modes
Our focus here is on the zero eigenvalue,  and hence on the zero mode of $\hat{H}_0$. In this case there are  the two independent zero modes 
\begin{equation}
\Psi_1 = \sqrt{\phi};\quad  \Psi_2 = \sqrt{\phi}\log \phi.
\end{equation}
%{\color{red} Although, the boundary conditions for the domain $\phi(a)\geq \Lambda$ can be given by any unitary matrix $U$ defined
%in the space of square integrable functions on the surface $\phi(a)=\Lambda$ \cite{Asorey:2004kk}, the only ones that can be renormalized to the whole space  $\phi(a)<\infty$ are given by the simple choice of an angular parameter $\theta\in(0,2\pi)$ which selects a {\it radial} asymptotic zero mode
%\begin{eqnarray} \Psi_\theta=\cos \theta\  \Psi_1+\sin \theta\  \Psi_2 \end{eqnarray}
%and fix the boundary conditions of  $\hat{H}_0$ by the asymptotic boundary condition
%\begin{equation}\lim_{\Lambda\to 0}\left(\Psi_\theta\partial_\phi \Psi -  \Psi \partial_\phi \Psi_\theta\right)=0.
%\label{fbc}
%\end{equation}}
Notice the asymptotic zero modes corresponding to the higher modes $\Delta_\Lambda$ vanish
at the singular points where $\phi(a)$ vanishes and do not require extra parameters to fix the
boundary condition. In some sense the simple structure of the singularity simplifies the
analysis of the boundary conditions \cite{{AS:2013}}.

We can understand the origin of zero modes very simply before the transformation (\ref{H0}): they are just the constant function and $\log \phi$. They correspond to the $\sqrt r$ and the $\sqrt r \log r$  zero modes of (\ref{eqn_1.3}) for $l=0$ and $d=2$.

From the boundary condition  it follows that the functions in the domain of
the Hamiltonian $\hat{H}_0$ have an  asymptotic behavior similar of $\Psi_\theta$, i.e.
\begin{equation}\label{eqn3.10}
\Psi \sim \cos \theta\  \Psi_1+\sin \theta\  \Psi_2  \quad \textrm{as  } \phi(a) \rightarrow 0. 
\end{equation}

In fact it is the relative coefficient $\tan \theta$   (which can also be infinite) between $\Psi_1$ and $\Psi_2$  is what matters.

Equation (\ref{eqn3.10}) refers only to the behaviour of $\Psi$ as $\phi(a) \rightarrow 0$. For large $a_i$ and/or large $\phi(a)$,  square integrability requires that $\Psi \rightarrow 0$. For example, the wave function  %$\Psi$ can be of the form
\begin{equation}
\Psi = ( \Psi_1 + \tan\theta \Psi_2) e^{-\phi(a )\sum_{i=1}^3 a_i^2}. 
\end{equation}
which is globally defined for any $a$ with $\phi(a)>0$ belongs also to the domain 
of the Hamiltonian $\hat{H}_0$.

{
As emphasised, these zero modes appear naturally in the theory and there is nothing mysterious about them. Working with the volume form $\phi(a) \prod_{i} da_i $ instead of $ \prod_{i} da_i $,  $\Psi_1$ corresponds to a constant  zero mode and of course, it is a zero mode of the Laplacian.  The numerical computation to glueball masses in \cite{Acharyya:2016fcn}  corresponds to a variational calculation around this mode which then brings in also non-constant functions.  Therefore, that is the $\theta =0$ case.

When nothing special happens at the $\phi (a) =0$ boundary,  we only consider $\Psi_1$. Instead if we consider  both $\Psi_1$ and $\Psi_2$  with a nonzero value $\theta$, there can be nontrivial physical effects at the $\phi(a)=0$ boundary,  like emergence of edge localized states. We demonstrate these in the following section.

\section{Edge States: Numerical Results}

In this section, we {calculate an upper bound of} the  ground state energy of 
\begin{equation}
\mathcal{H} = \hat{H}_0 + V(M)
\end{equation}
(where $\hat{H}_0$ and $V(M)$ are defined in (\ref{eqn2.17}) and (\ref{pot}) respectively)  using variational calculation and show the existence of edge states as described in the previous section.  

For certain choices of $\theta$  to be discussed below, energy becomes negative making the system unstable. Such $\theta$ have to be rejected,  without further physical inputs.
 Of course, as we are doing only a variational calculation, we can only numerically check that the energy is  positive. That carries the uncertainties of numerical estimates.

As a variational ansatz for the ground state, we consider 
\begin{equation}
\zeta_k =  (\Psi_1 +\tan\theta \Psi_2) e^{-k\phi(a )\sum_{i=1}^3 a_i^2}, \quad\quad k>0
\end{equation}
which has the asymptotic behaviour (\ref{eqn3.10}) and is square integrable in $a_1\geq a_2 \geq a_3 \geq a_3\geq0$.   Here, $k$ is the variational parameter. 

For fixed values of $\tan\theta $, we numerically compute 
\begin{eqnarray} \label{gs1}
E(k) = \frac{  \displaystyle\int_0^\infty da_1\int_0^{a_1}  da_2\int_0^{a_2}da_3 \ \zeta_k^\dagger (\mathcal{H}  \zeta_k) }{ \displaystyle \int_0^\infty da_1\int_0^{a_1}da_2  \int_0^{a_2} da_3 \ \zeta_k^\dagger \, \zeta_k }
\end{eqnarray}
as a function of the variational parameter $k$. {This provide us by the Rayleigh-Ritz theorem an upper bound of  the ground state energy $E$ of the system}:
\begin{equation}\label{E_defn}
E<E({k_0}), \quad \textrm{where }\quad \frac{d E(k)}{dk}\Big|_{k=k_0}=0. 
\end{equation}

We also evaluate 
\begin{eqnarray}
E_0(k) =\frac{ \displaystyle \int_0^\infty da_1\int_0^{a_1}  da_2\int_0^{a_2}da_3\  \zeta_k^\dagger (\hat{H}_0 \zeta_k) }{  \displaystyle \int_0^\infty da_1\int_0^{a_1}da_2  \int_0^{a_2} da_3\ \zeta_k^\dagger\,  \zeta_k }
\end{eqnarray}
and show that there exist certain choices of $\tan\theta $ for which $E_0 (k_0) <0$.

Near $\phi(a)=0$, as both $\zeta_k \rightarrow 0 $  and $\mathcal{H}  \zeta_k \rightarrow 0$, the contribution to the integrals in  (\ref{gs1}) is very small from this region. Consequently, we can deform $E(k)$, 
\begin{eqnarray} \label{gs2}
E(k) = \displaystyle\frac{ \displaystyle \int_{2\epsilon}^\infty  da_1\int_{\epsilon} ^{a_1-\epsilon}  da_2\int_{0} ^{a_2-\epsilon}da_3  \zeta_k^\dagger (\mathcal{H}  \zeta_k) }{ \displaystyle \int_{2\epsilon}^\infty da_1\int_{\epsilon} ^{a_1-\epsilon}  da_2\int_{0} ^{a_2-\epsilon}da_3  \zeta_k^\dagger  \zeta_k },
\end{eqnarray}
where $\epsilon$ is very small. 
This is done for the following reason. The integral in the numerator involves $\partial_i \phi(a)$ which does not have the same zeros as $\phi (a)$ and $\partial_i \phi(a)$  can be large even near $\phi(a)=0$. The integral involves product of $\phi (a)$ and $\partial_i \phi(a)$'s and numerical evaluation  might be erroneous  because of multiplying a very small number (the zero of the computer) with a very large number .  

We checked that as $\epsilon$ is reduced,  the integrals converge and that it is enough to consider $\epsilon=0.01$ for a good estimation of $E(k)$.

%\red{\underline{NOTE}: Please change in all graphics  $\frac{\beta}{\alpha}$ by $\tan \theta\,$}

For $\tan \theta=0$, $E_0(k)$ is monotonic and positive, while for  $\cot \theta=0$, $E_0(k)$ is monotonic and negative. 
Consequently, when we choose $0<|\theta| \ll \frac{\pi}{2}$,  the energy functional $E(k)$ can be positive and have a minima. On the other hand, for $0\ll|\theta|<\frac{\pi}{2}$, there might be minima, but $E(k)$ might be negative.

In the following, we study the two regimes separately. 

\subsection*{\textit{$0<|\theta| \ll \frac{\pi}{2}$}:}
First, we consider $0<|\theta| \ll \frac{\pi}{2}$ and estimate $E_0(k)$ and $E(k)$ numerically. 
For various values of $ \theta$ and $g$, we have plotted $E(k)$  as a function of $k$ in Fig. \ref{fig1}. 
\begin{figure}[h]
\begin{center}
\includegraphics[scale=0.7]{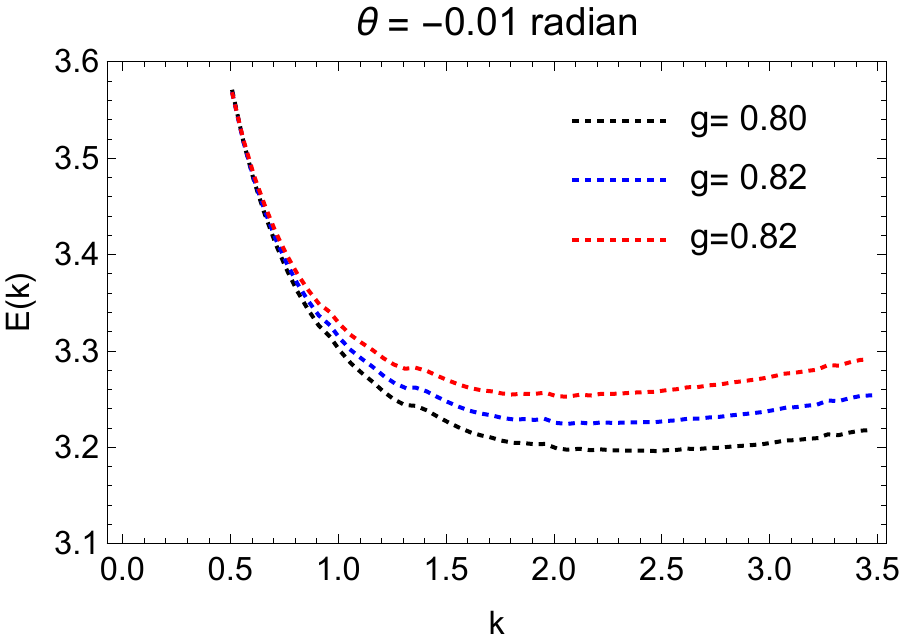} \quad\quad\quad
\includegraphics[scale=0.7]{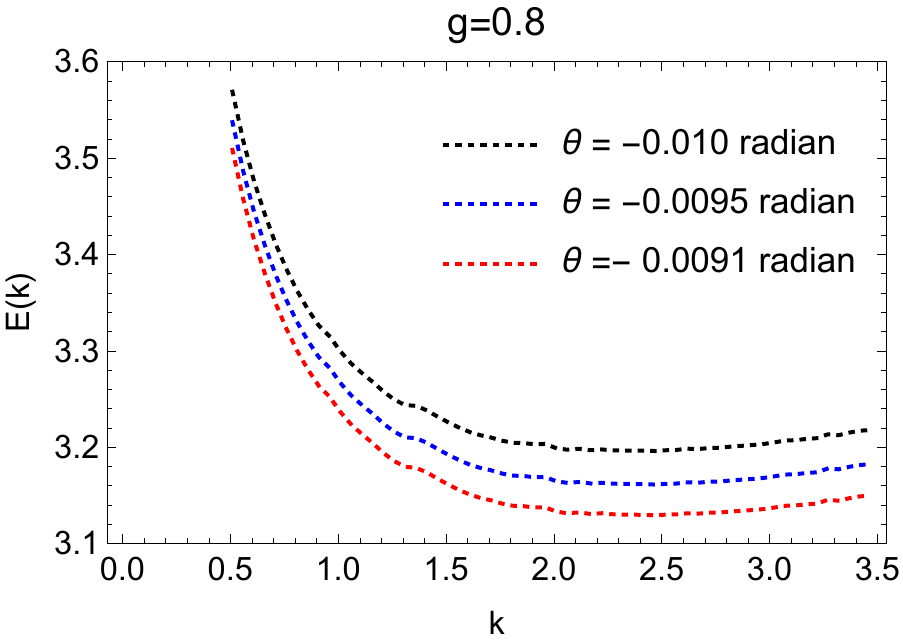}
\caption{Plot of $E(k)$ as a function of $k$ for various values of $g$ with a fixed $\theta$ (left) and for various values of $\theta$ with a fixed  $g$ (right). }\label{fig1}
\end{center}
\end{figure}
%\begin{figure}[h]
%\begin{center}
%\includegraphics[scale=0.7]{fig1a_pos.pdf}
%\caption{Plot of $E_k$ as a function of $k$ for various values of $\theta$ and fixed value of $g=0.8$ . }\label{fig1a}
%\end{center}
%\end{figure}
The minima of $E(k)$ (as in the plots) give upper bounds of the total energy  $E(k_0)$, which are positive such choices of $\theta$.  For various values of  $\theta $,  $E(k_0)$ as a function of the coupling constant $g$ is shown in  Fig. \ref{fig3}.   
%\begin{figure}
%\begin{center}
%\includegraphics[scale=0.6]{figke1_pos.pdf}
%\caption{Plot of $E_0(k0)$  as a function of $g$. }\label{fig2}
%\end{center}
%\end{figure}
\begin{figure}
\begin{center}
\includegraphics[scale=0.8]{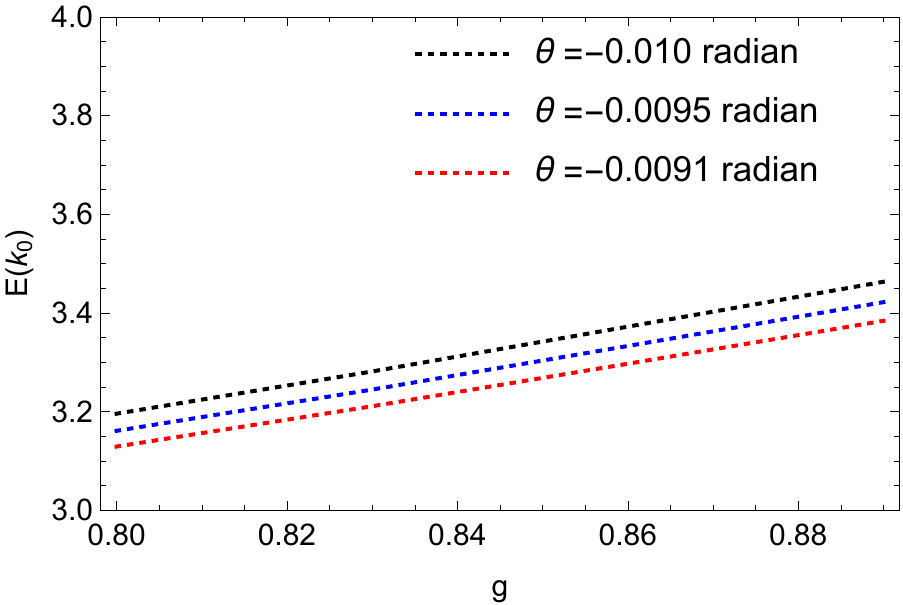}
\caption{Plot of $E(k_0)$  as a function of $g$. }\label{fig3}
\end{center}
\end{figure}

%In figure 4, we show that for the above choices of $\tan\theta$, $E_0(k_0)$ is negative for all $g$ and hence $\zeta_{k_0}$ are edge states. 
%\begin{figure}[h]
%\begin{center}
%\includegraphics[scale=0.7]{Figure4.pdf}
%\caption{Plot of $E_0(k_0)$  as a function of $g$. }
%\end{center}
%\end{figure}
For such choices of $\theta$,  due to the exponential factor in the modes $\zeta_{k_0}$, they might localized near the $\phi(a)=0$ boundary. That can be demonstrated by plotting $|\zeta_{k_0}|^2 (a_1,a_2, a_3) $ as a function of $a_1, a_2 $ and $a_3$ for fixed $ \theta\,$ and $g$. In Fig. \ref{fig4},  we  have shown the contour plots of $|\zeta_{k_0}|^2 (a_1,a_2, a_3) $ for fixed $a_1$'s in the range $a_2\in [0,a_1]$ and $a_3 \in [0,a_2]$. The darker regions denote higher values of  $|\zeta_{k_0}|^2$. 
\begin{figure}[t!]
\begin{center}
\includegraphics[scale=0.5]{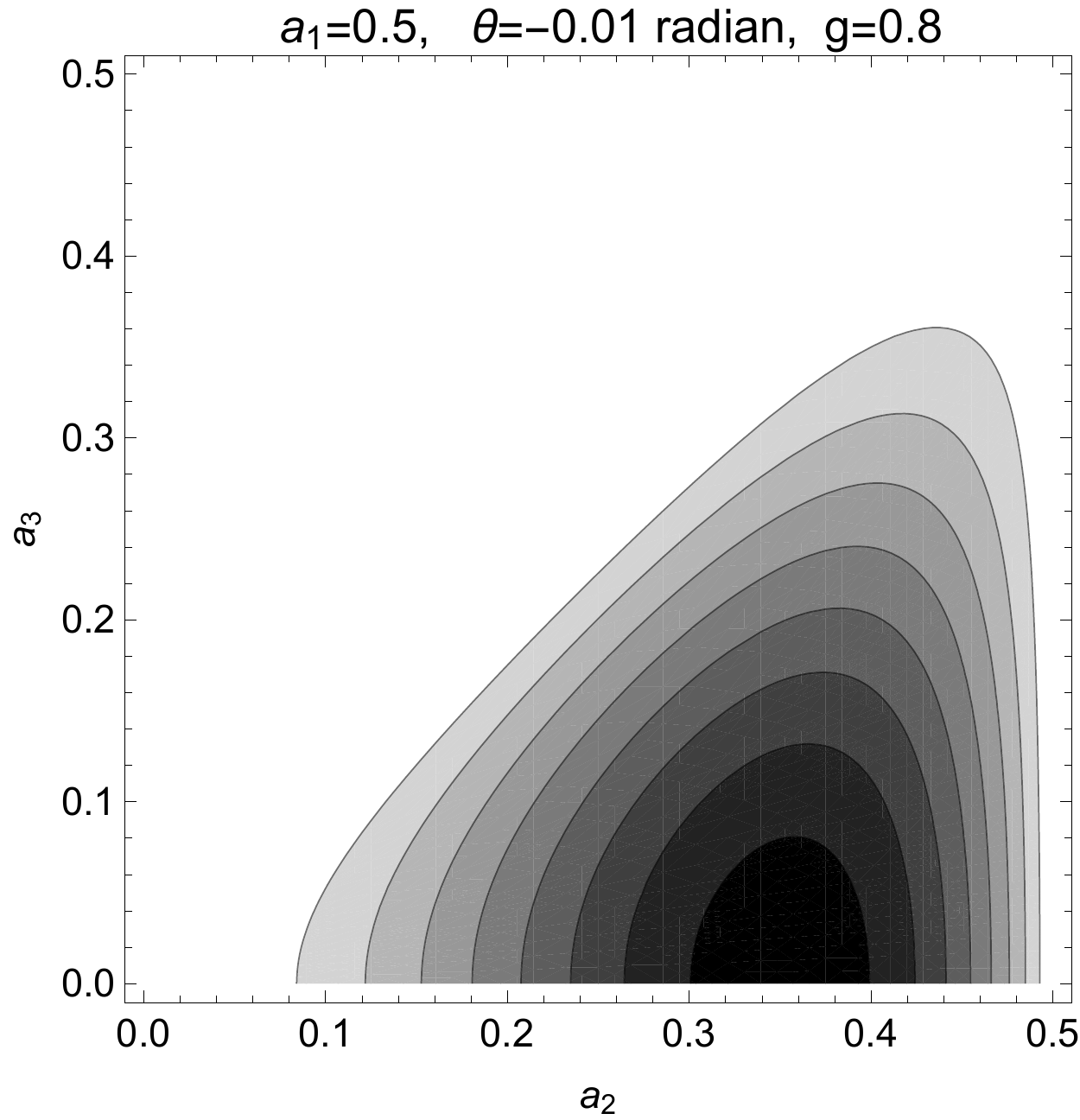}
\includegraphics[scale=0.5]{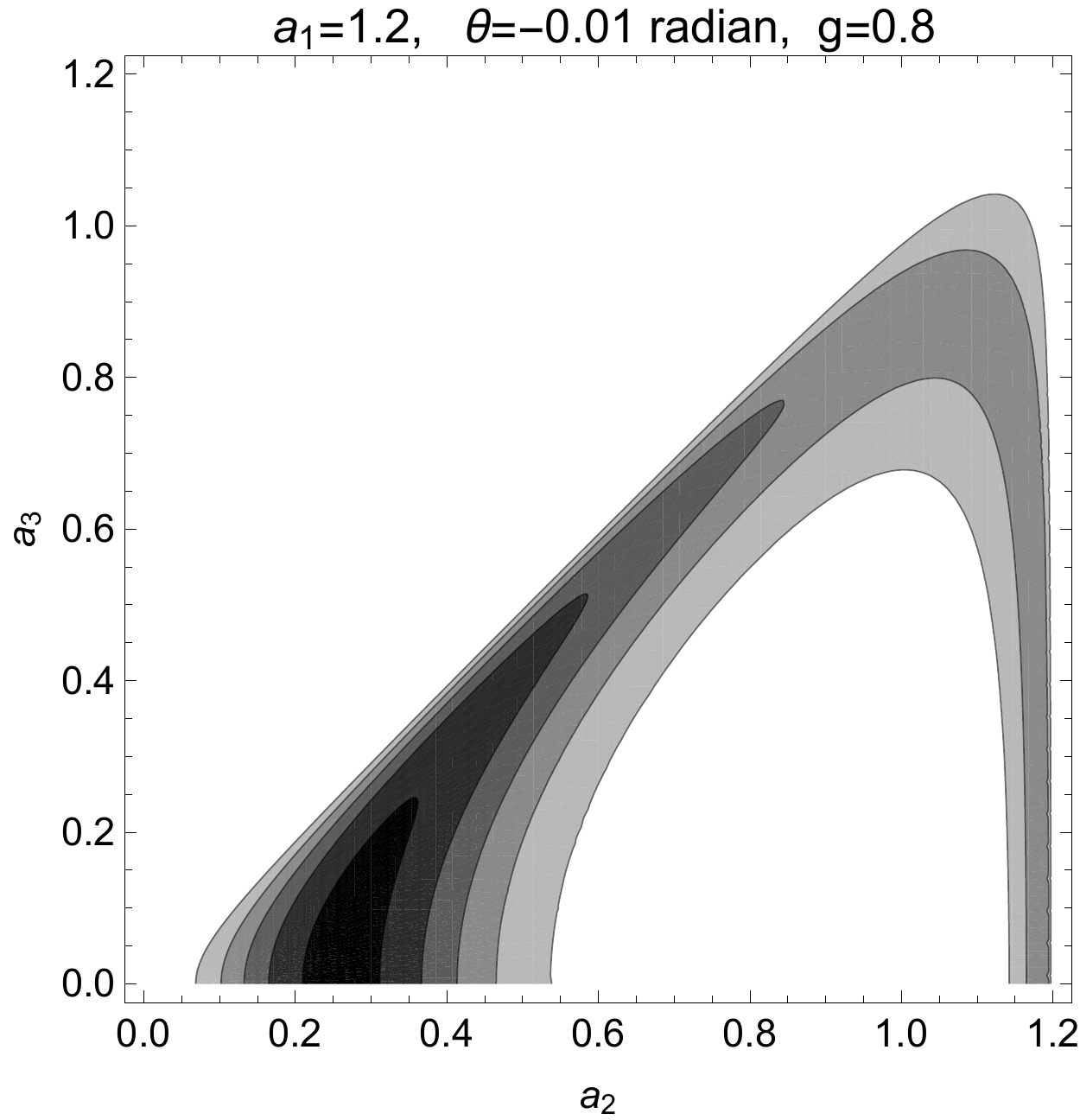}\vspace*{.3cm}
\includegraphics[scale=0.5]{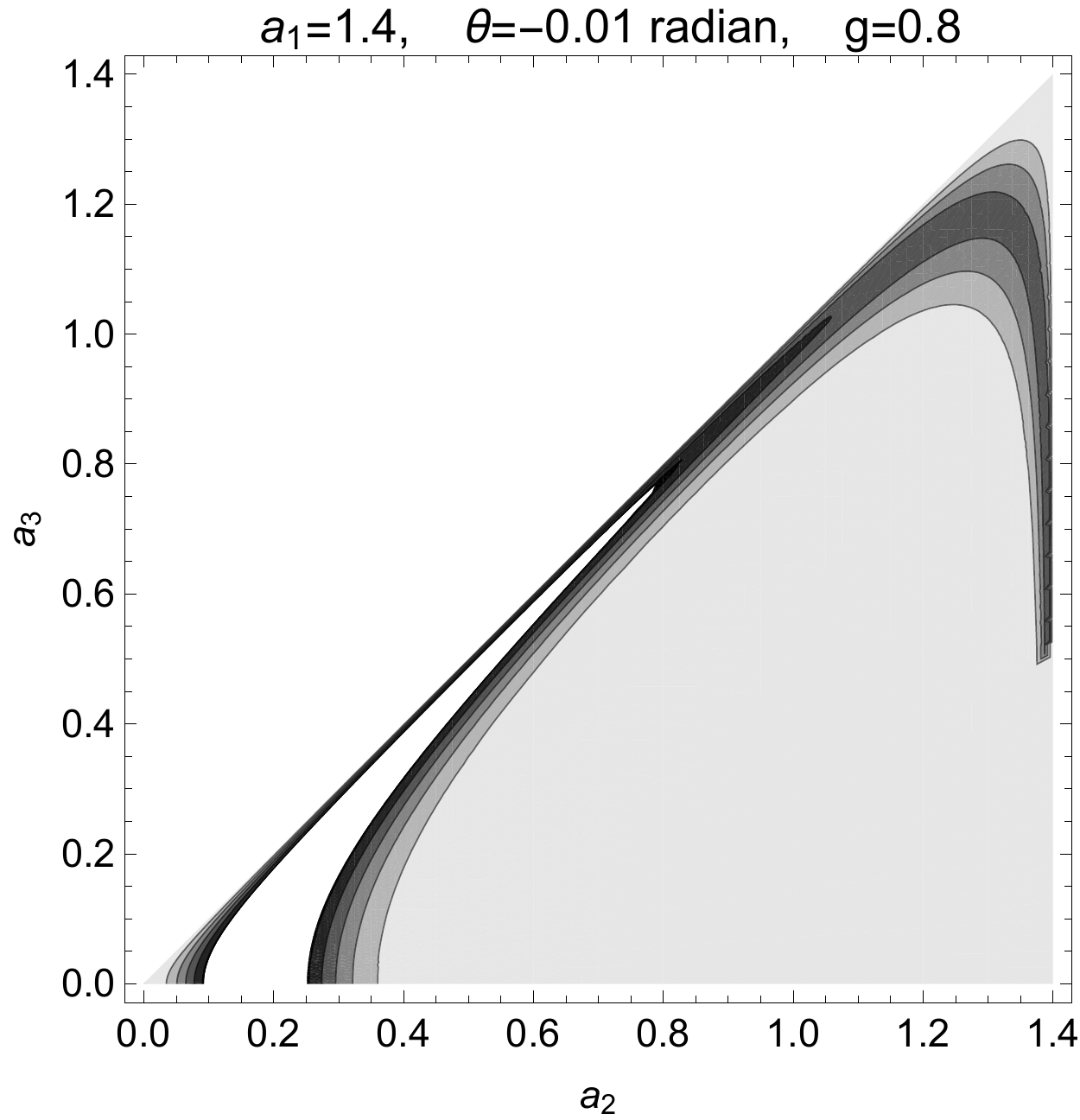}
\includegraphics[scale=0.5]{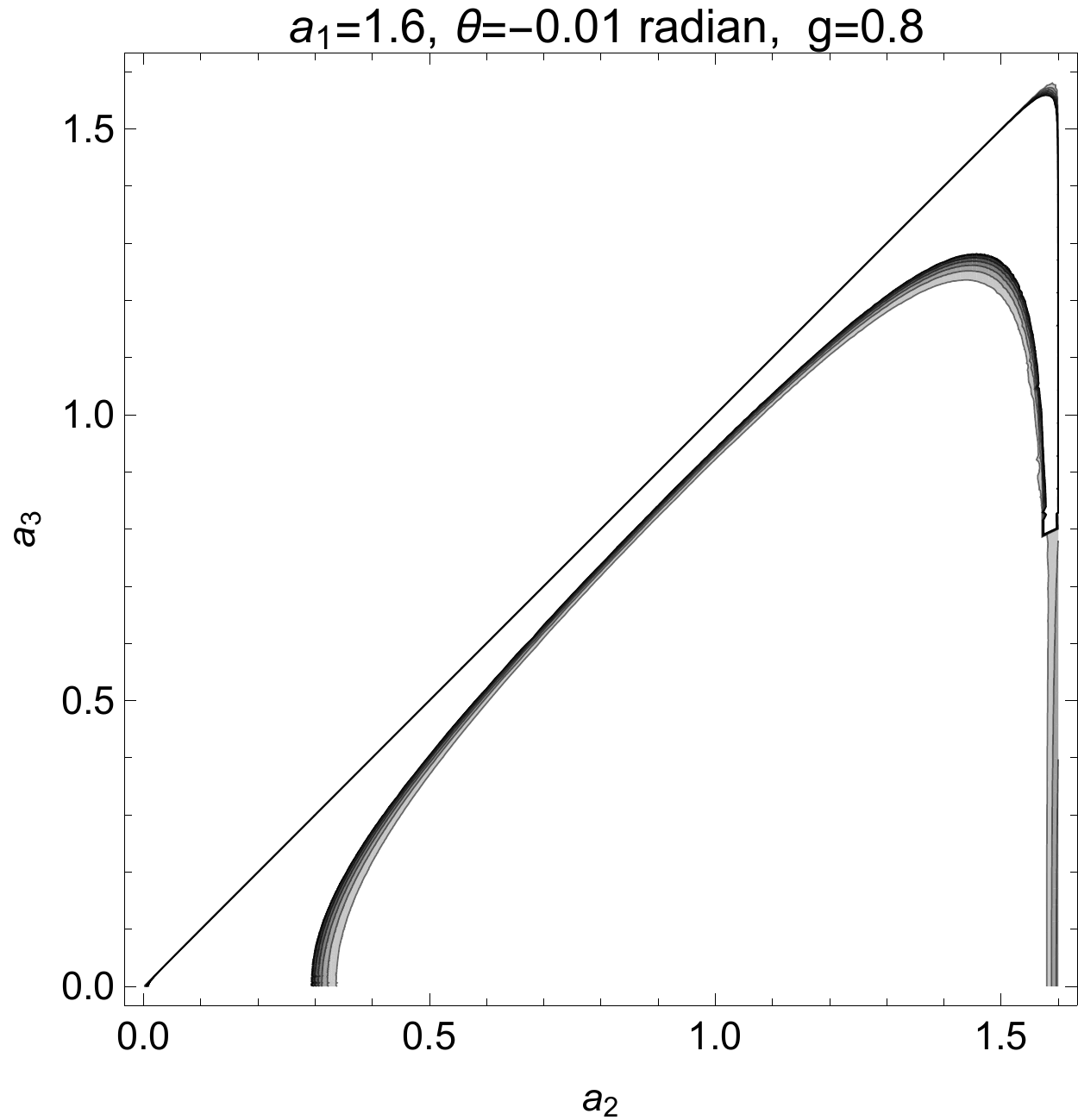}
\caption{Plot of $|\zeta_{k_0}|^2$  as a function of $a_2$ and $a_3$ for fixed values of $a_1$ and $\theta\,=-0.01$ radian. Here, we used $g=0.8$ and $k_0 \approx 2.35$ which is obtained from the minima in Fig. \ref{fig1} with $\theta\,=-0.01$ radian.  }\label{fig4}
\end{center}
\end{figure}

From the figures, we can see that when $a_1$ is small, $a_1\geq a_2 \geq a_3 \geq 0$ is the region close to the vertex of the the wedge-like region spanned by $a_1$, $ a_2$ and $a_3$. We see that $|\zeta_{k_0}|^2$ is significantly large in this region. For larger $a_1$,  $|\zeta_{k_0}|^2$ is localised near the $a_2=a_3$ boundary (the diagonal lines in the boxes), $a_1=a_2$ boundary (the right hand limits of the boxes) and near $a_1=a_2=a_3$ boundary (the top-right corners of the boxes), while in the interior region where the $a_i$'s are distinctly different, $|\zeta_{k_0}|^2$ is significantly damped.  Thus we conclude that $\zeta_{k_0}$ describes  states localised near the $\phi(a)=0$ boundary. 

 As $|\theta|$ decreases, the value of $k_0$ decreases which can be seen from Fig. \ref{fig1}. As a result, the exponential factor decays slowly for smaller $|\theta|$ and $|\zeta_{k_0}|^2$ spreads more into the bulk.  This is consistent with the fact that there is no edge state at $\theta =0$.

%We can now summarise the positive energy edge localised glue ball states. They occur for $0< |\theta|\ll \pi/2$ and have mass values $\sim 3-4$. Fig. \ref{fig3} shows the masses of these glue balls for various values of $\theta$ and $g$. 

\subsection*{\textit{$0\ll |\theta| < \frac{\pi}{2}$}:}
For a large value of $|\theta|$, we indeed find that $E(k)$ is negative. For various values of $ \theta$ and $g$, we have plotted $E(k)$  as a function of $k$ in Fig. \ref{fig5}. 
\begin{figure}[h]
\begin{center}
\includegraphics[scale=0.7]{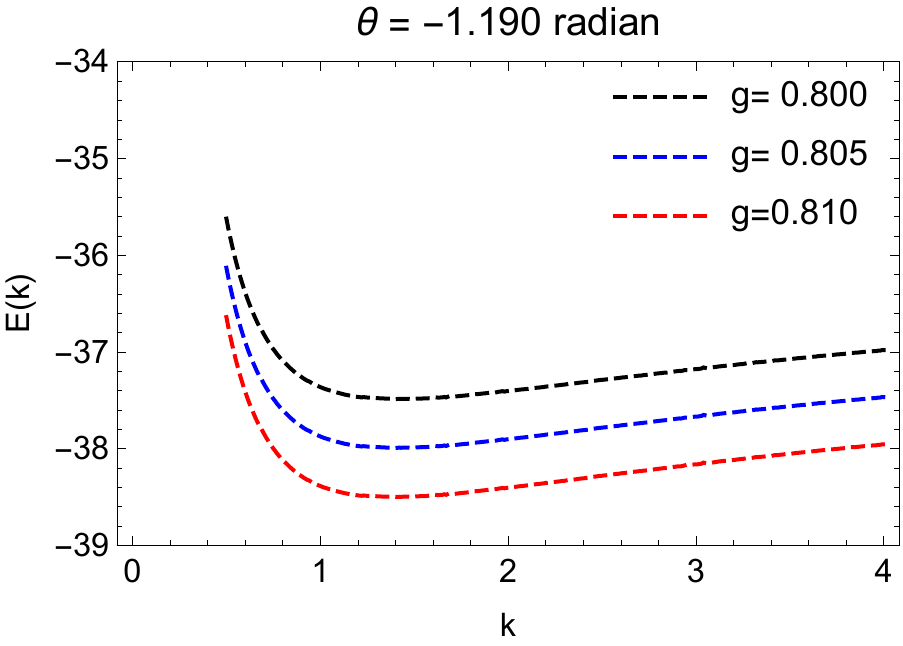} \quad\quad \quad 
\includegraphics[scale=0.7]{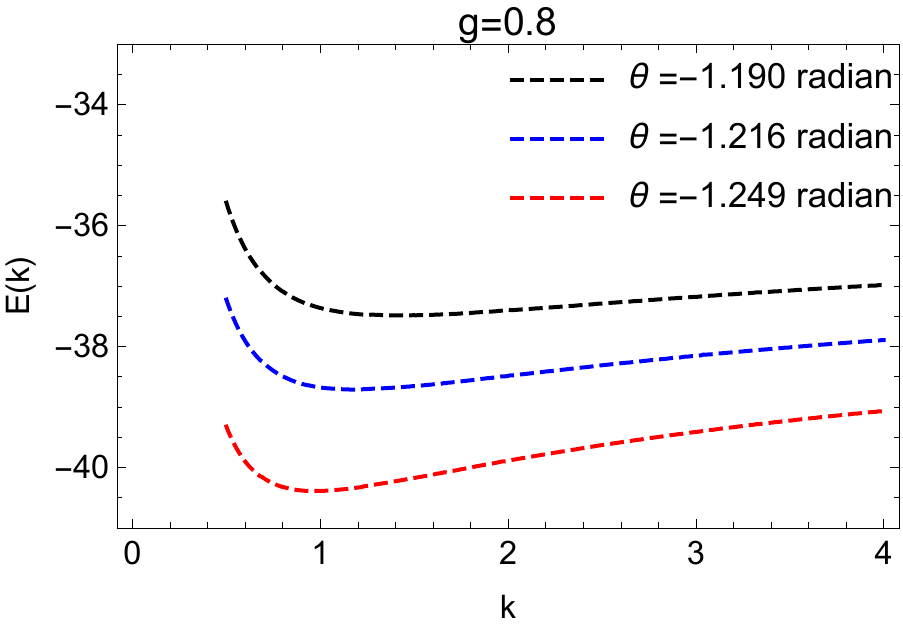}
\caption{Plot of $E(k)$ as a function of $k$ for various values of $g$ with a fixed $\theta$ (left) and for various values of $\theta$ with a fixed  $g$ (right). }\label{fig5}
\end{center}
\end{figure}
%\begin{figure}[h]
%\begin{center}
%\includegraphics[scale=0.7]{fig1a_neg.pdf}
%\caption{Plot of $E_k$ as a function of $k$ for various values of $\theta$ and fixed value of $g$ (left) and for various . }\label{fig5a}
%\end{center}
%\end{figure}

Again, the minima gives the upper bound of the total energy $E(k_0)$. We have plotted  $E(k_0)$ as a function of the coupling constant $g$ for various values of  $\theta $  in  Fig. \ref{fig7}.   
%\begin{figure}
%\begin{center}
%\includegraphics[scale=0.7]{figke1_neg.pdf}
%\caption{Plot of $E_0(k0)$  as a function of $g$. }\label{fig6}
%\end{center}
%\end{figure}
\begin{figure}[t!]
\begin{center}
\includegraphics[scale=0.8]{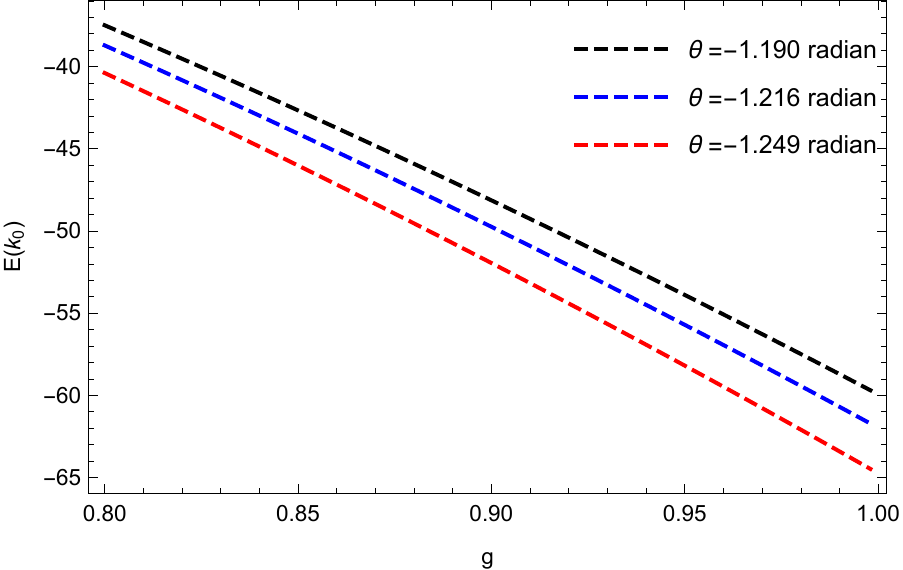}
\caption{Plot of $E(k_0)$  as a function of $g$. }\label{fig7}
\end{center}
\end{figure}

Again, if we plot $|\zeta_{k_0}|^ 2$ as a function of $a_1, a_2$ and $a_3$ (as in Fig. \ref{fig4})  for a fixed $\theta$ and $g$, we will find that the modes are localised near $\phi=0$ boundary.  So these are edge states but with negative { $\langle H_0+V(M)\rangle$.}

As $|\theta|$ increases towards $\frac{\pi}{2}$,  $k_0$ decreases (as can be seen from Fig. \ref{fig5}). Consequently, the modes spread more into the bulk. 

{ When $0> \langle H_0+V(M)\rangle \geq - C(R)$, the total energy of these edge states are positive and such edge states are physical. 

On the other hand, when   $ \langle H_0+V(M)\rangle < - C(R)$, the total energy is negative. }
As glueballs are bosons, if such negative energy states exist, there is no Pauli exclusion principle to prevent states with arbitrary number of such edge localised glueballs with negative energies. Therefore, the energy will be unbounded from below  and the vacuum will be unstable.  Therefore, these edge states should be considered \textit{unphysical}.

%We can now summarise the positive energy edge localised glue ball states. They occur for $0< |\theta|\ll \pi/2$ and have mass values $\sim 3-4$. Fig. \ref{fig3} shows the masses of these glue balls for various values of $\theta$ and $g$. 
\section{Negative Energy Edge states and{ Indications of  Phase Transition} }

%\red{Since $E_0(k_0)<0$ there are physical states where $E_0<0$. We also have shown that they are edge states.  The presence of edge states is simply due to the fact that  $\hat{H}_0$ have negative energies. However, the
%total energy $E$ of edge states with respect to the  Hamiltonian $\hat{H}=\hat{H}_0+V(M)$ 
%can be positive which save the consistency of the theory because since  
%gluons are bosons, there is no Pauli exclusion principle to prevent states with arbitrary number of these edge state 
%glueballs with negative energies. The energy thus would  be unbounded below and the vacuum unstable. However,
%%this does not occurs for small values of $\theta<<1$.}
%
%\red{\underline{NOTE}:From here on changes are needed along the last Nirmalendu remarks based on numerical results that
%show that the upper bound of the total energy E might be positive in spite of the fact that contribution of the 
%kinetic term is negative. Notice that  the virial theorem cannot be used because we are breaking conformal
%invariance by the introduction of a non-trivial boundary condition}

%That is often the case also for $\hat{H}=\hat{H}_0+V(M)$. 
%Since, gluons are bosons, there is no Pauli exclusion principle to prevent states with arbitrary number of these glueballs. The energy thus can be unbounded below.  

{ In the previous section,  we have shown that the total energy of the edge states in the matrix model can be negative {(i.e. $ \langle H_0+V(M)\rangle < - C(R)$)}. In a pure gauge theory, these negative energy states are unphysical. However, as we argue below, on inclusion of matter fields, these states can have positive energy and therefore, can exist. }

 A similar situation has been treated by Asorey et. al. \cite{Asorey1,Asorey2} for spin zero (and one) fields on  a spatial disk $\mathcal{D}$ with boundary $\partial\mathcal{D}$.  If $\hat{n}$ is the outward-drawn unit normal at $\partial \mathcal{D}$ and $\partial_n$  denotes $\hat{n}\cdot \vec{\nabla}$ at $\partial \mathcal{D}$, the scalar Laplacian $\Delta = -\sum_i \frac{\partial^2\,\,}{\partial x_i^2} $ is (essentially self-adjoint for the Robin boundary conditions
\begin{eqnarray}\label{eqn4.1}
(\Psi+i\partial_n \Psi)\Big|_{\partial \mathcal{D}} = e^{i \theta} (\Psi-i\partial_n \Psi)\Big|_{\partial \mathcal{D}}, \quad\quad e^{i \tilde{\theta}} \in U(1). 
\end{eqnarray}
When $e^{i\tilde{\theta}} =1$, (\ref{eqn4.1}) gives the Neumann boundary condition $\partial_n \Psi \Big|_{\partial \mathcal{D}}=0$, while if  $e^{i\tilde{\theta}} =-1$, it gives the Dirichlet boundary condition $ \Psi \Big|_{\partial \mathcal{D}}=0$. Near Dirichlet point, there are Robin boundary conditions 
\begin{equation}\label{eqn4.2}
\Psi\Big|_{\partial \mathcal{D}}= \lambda\partial_n \Psi\Big|_{\partial \mathcal{D}}.
\end{equation}
It is an important result of \cite{Asorey1} that the Laplacian $-\sum_i \frac{\partial^2\,\,}{\partial x_i^2} $ has edge-localised negative energy states if $\lambda>0$. Hence, the free  Laplacian  $-\sum_i \frac{\partial^2\,\,}{\partial x_i^2} $ cannot be second quantised. 

But it was proved \cite{Asorey2} that  $-\sum_i \frac{\partial^2\,\,}{\partial x_i^2} $ has a lower bound:
\begin{equation}
 -\sum_i \frac{\partial^2\,\,}{\partial x_i^2} \geq -m_0^2(\lambda).
\end{equation}
Hence,
\begin{equation}
 -\sum_i \frac{\partial^2\,\,}{\partial x_i^2}+m_0^2(\lambda) \geq 0,
\end{equation}
and the Lagrangian density
\begin{equation}\label{eqn4.5}
\mathcal{L}= -\phi^\ast \left(\frac{\partial^2\,\,}{\partial x_0^2}-\sum_i \frac{\partial^2\,\,}{\partial x_i^2}-m_0^2(\lambda)\right)\phi
\end{equation}
allows a consistent quantisation.

In addition, (\ref{eqn4.5}) is the Lagrangian for a superconductor. (For the latter, $\phi$ should be a vector field, but that is not important here). Further the field $\phi$ in the ground state decays as it enters $\mathcal{D}$ due to Meissner effect: 
\begin{equation}\label{eqn4.6}
\phi=\phi_0 e^{-(r_0- r)m_0^2(\lambda)}
\end{equation}
near $\partial \mathcal{D}$, $r$ decreasing away from the boundary.  From (\ref{eqn4.6}), we get 
\begin{equation}
\left(\phi-m_0^2(\lambda)\partial_r \phi\right)\Big|_{\partial\mathcal{D}}=0. 
\end{equation}
This is (\ref{eqn4.2}) for $\lambda = m_0^2 (\lambda)$.

Thus the negative energy levels signify the transition to the superconducting phase. 

The wave functions $\Psi$ vanishing at $\partial\mathcal{D}$ have  nonnegative energies even without $m_0^2(\lambda)$:
\begin{equation}
\left(\Psi,  -\sum_i \frac{\partial^2\,\,}{\partial x_i^2} \Psi\right) =\left (  \sum_i \frac{\partial\,\,}{\partial x_i}\Psi,  \sum_i \frac{\partial\,\,}{\partial x_i}\Psi\right) \geq 0 \quad \textrm{for }  \Psi\Big|_{\partial\mathcal{D}}=0.
\end{equation}
Hence, with the addition of $m_0^2(\lambda) $, the edge states get lifted to positive energies (which can be adjusted to be low-lying), while bulk states get gapped, with bulk energies $> |m_0(\lambda)|$.

The possibility of  ``superconducting''  phases have been considered in the quark-gluon plasma phase of QCD \cite{sanatan1, sanatan2}.  {Such color superconductivity is expected to be the ground state when the temperatures are low and the baryon chemical potential is high. When massless quarks are coupled to the pure Yang-Mills theory, indeed color-flavour locked phase or 2SC (when one quark does not participate in the condensation) phases can emerge \cite{bailin, Iwasaki, Alford}. The  superconducting phases emerge when the global symmetries $SU(3)_F$ and $U(1)_B$ are broken.
In the quark-gluon plasma phase, the symmetry group is $\frac{SU(3)_C \times SU(3)_F \times U(1)_B}{Z_3\times Z_3}$.   In the superconducting phase, the pairing of two quarks of same helicity is dominant and the presence of this diquark condensate spontaneously breaks the symmetry to $\frac{SU(3) \times Z_3}{Z_3\times Z_3}$. This spontaneous breaking of the flavour symmetry and $U(1)_B$ naturally leads to a phase of massive gluons. 

In the matrix model too, we can consider gluons coupled to the flavour symmetry breaking diquark condensate. In that case, the matrix model is constructed by pulling back the Maurer-Cartan one-forms of   $\frac{SU(3) \times Z_3}{Z_3\times Z_3}$ instead of $\frac{SU(3)_C \times SU(3)_F \times U(1)_B}{Z_3\times Z_3}$. In this matrix model, the gluons are massive. The mass term lifts the edge levels to positive energies and at the same time creates a gap in the bulk levels making them incompressible.  Thus, in such a massive gluon phase, the aforementioned edge states do exist.  }

%Work on the nature of the potential $V(M)$ is in progress \cite{vaidya_prep}. 

\section{Edge States:  Angular Momentum and Colour} 

The Schr\" {o}dinger  Hamiltonian on $\mathbb{R}^3$, on separation of variables, acquires the centrifugal term $\frac{l(l+1)}{r^2}$. This term eliminates the boundary condition ambiguities at $r=0$ from all except the S-wave. 

But it is a surprising result of Iwai (sections 3.3, 5.2 5.3 in \cite{iwai})\footnote{This point was emphasised to us by Sachindeo Vaidya.}  that the induced potential in the Hamiltonian $H=\Delta+V(M)$ (see (\ref{eqn_delta})) for colour or angular momentum states is \textit{finite} as $\phi(a)\rightarrow 0$. That means that edge states are present also with angular momentum and colour excitations.

Their energies will depend on angular momentum and colour because the induced potential depends on them. It will be interesting to study this energy dependence on angular momentum and colour.

\section{Discussions} 
In a matrix model of $SU(2)$ Yang-Mills theory, the Hamiltonian requires boundary conditions on the boundaries of $\textrm{Mat}_3 (\mathbb{R})$. We have shown that for certain choices of these boundary conditions, there are glueball states localized near the boundaries.  The energy of these edge states can be negative, in which case they can only be physical, if, a London-like mass term is added to lift the total energies to a positive values.  In presence of matter, such a mass term can indeed be generated and there, these edge states comprised of massive gluons constitute a superconducting phase of QCD.  

{ In this matrix model, one can construct colored states of the Hamiltonian as well. However, as shown in \cite{Balachandran:2014voa, mm_bsa}, all observables are color singlet functions of $M$. Thus, the colored states naturally decouple from the color singlet theory. 

Also, the colored states are mixed, while the colorless ones are pure \cite{Balachandran:2014voa, mm_bsa}. That is why it is not possible to evolve from a colourless state to the tensor product of colored states by unitary time evolution. }

Under $M_{ia} \rightarrow  -M_{ia}$, the singular values  are invariant: $a_i \rightarrow a_i$. Consequently, the ground state obtained by the variational computation has even parity, as expected.  {Here, we used the zero modes of $H_0$ to construct the variational ansatz. For a better approximation, we can, in principle, include non-zero modes as well in the variational ansatz. However, there will be additional computational complexity owing to many non-vanishing terms.}  

Although, we have demonstrated the presence of these edge states in a $SU(2)$ Yang-Mills theory, the analysis and the conclusions can be readily extended to $N>2$.  However, for large $N$, the singular value decomposition becomes difficult because under color, the gauge fields transforms as  $M \rightarrow M (Ad(h))^T$, $h \in SU(N)$.   

To study the large $N$ limit, we should start with the observation: %Despite that we can study large $N$ limit inspired by many works on string theory.
our matrix model  is very similar to a three-matrix model describing $N$-coincident D-branes coupled to a  Ramond-Ramond 4-form field \cite{Myers:1999ps}. 
In particular, the potential of our matrix model 
\begin{eqnarray}
V(M) =  \frac{1}{2  R g^2}Tr\left( M_iM_i +i \epsilon_{ijk} M_i[M_j, M_k]-\frac{1}{2}[M_i,M_j]^2  \right), \quad\quad M_i \equiv M_{ia}T_a, \label{full_pot}
\end{eqnarray}
($T_a$'s are generators of $SU(N)$ in the fundamental representation) has extrema describing $N\times N$ fuzzy sphere algebras (similar to \cite{Myers:1999ps}).   Here, the difference between $N=2$ and $N> 2$ appears: for $N=2$, only the non-trivial extremum is described by the fuzzy sphere algebra in a 2-dimensional irreducible representation, while for $N>2$, the algebra can be  $N$-dimensional irreducible  or  any possible reducible representation of $SU(2)$.  

{The vacua corresponding to the irreducible  and the reducible representations are degenerate and transitions between them occur by quantum tunneling (as discussed in \cite{Jatkar:2001uh}), which we intend to study in future. 

%As in \cite{Myers:1999ps}, in our matrix model (for $N>2$) too, the reducible representations have higher energy.  However, unlike \cite{Myers:1999ps}, in our matrix model, due to presence of the ``mass term'' (the first term in the r.h.s. of (\ref{full_pot})), it is not possible to classically cascade down from the higher energy reducible representations to the lower energy irreducible one.  Such transitions  are only possible by quantum tunneling (as discussed in \cite{Jatkar:2001uh}), which we intend to study in future. 

One can also obtain the quantum states of these fuzzy spheres (both reducible and irreducible) by Gelfand-Naimark-Segal construction and compute the von Neumann entropy associated with the fuzzy spheres (as in \cite{Acharyya:2014nfa}) to study the evolution of the system. 

The setting in the previous paragraph is perfect to extend the study of the our matrix model to a large $N$ limit and discuss its possible equivalence to the Calogero model in the light of  \cite{Polychronakos:2001mi}.   This will be future direction of our investigation. }
 \\ \mbox{} \\\mbox{} \\
\textbf{Acknowledgements}\\
{ We are grateful to  Manuel Asorey for extensive collaboration and  his help in preparing the manuscript.}
 This work is  the continuation of the collaborative work done by A.P.B.with Sachin Vaidya and Amilcar Queiroz. He has benefited from many discussions with Sachin and also with Professor M.~S.~Narasimhan.
This work has been partially supported by the  grants FPA2015-65745-P (MINECO/FEDER)  and DGA-FSE 2015-E24/2

\end{document}